\documentclass[useAMS,usenatbib]{mn2e}
\usepackage[centertags]{amsmath}
\usepackage{amsfonts}
\usepackage{amssymb}
\usepackage{newlfont}
\usepackage{textcomp}
\usepackage{times}
\usepackage{graphicx}
\usepackage{subfigure}
\usepackage{psfrag}
\usepackage{amsmath}
\usepackage[at]{easylist}
\usepackage{bm}
\usepackage{rotating}

\usepackage{stmaryrd}

\def\reff@jnl#1{{\rm#1\/}}
\def\apj{\reff@jnl{ApJ}}       % Astrophysical Journal
\def\apjs{\reff@jnl{ApJS}}     % Astrophysical Journal, Supplement
\def\aaps{\reff@jnl{A\&AS}}    % Astronomy and Astrophysics, Supplement
\def\mnras{\reff@jnl{MNRAS}}   % Monthly Notices of the RAS
\def\prd{\reff@jnl{Phys.\ Rev.\ D}}    % Physical Review D

\newcommand{\beq}{\begin{equation}}
\newcommand{\eeq}{\end{equation}}

\newcommand{\be}{\begin{equation}}
\newcommand{\ee}{\end{equation}}
\newcommand{\aap} {A\&A}
\newcommand{\aapr} {A\&AR}
{\begin{enumerate}\setlength{\itemsep}{0mm}}%
{\end{enumerate}}
{\begin{enumerate}\setlength{\itemsep}{0mm}}%
{\end{enumerate}}

\def\tfrac#1#2{{\textstyle\frac{#1}{#2}}}
\def\vect#1{{\bmath{#1}}}

\title[PowellSnakes II: multi-frequency object detection]{Powellsnakes II:
a fast Bayesian approach to discrete object detection in multi-frequency astronomical data sets}
\author[P.~Carvalho et al.]
{Pedro Carvalho,$^1$\thanks{email: carvalho@mrao.cam.ac.uk}
 Gra\c{c}a Rocha,$^{2,3}$\thanks{email: graca@caltech.edu}
 M.P.~Hobson,$^1$\thanks{email: mph@mrao.cam.ac.uk}
 A.~Lasenby$^{1,4}$\thanks{email: a.n.lasenby@mrao.cam.ac.uk}
\\
\\
  $^1$ Astrophysics Group, Cavendish Laboratory, J.J.~Thomson Avenue, Cambridge CB3 0HE, U.K.\\
  $^2$ California Institute of Technology, Caltech, Pasadena, California, U.S.A.\\
  $^3$ Jet Propulsion Laboratory, JPL, 4800 Oak Grove Drive, Pasadena, California 91109\\
  $^4$ Kavli Institute for Cosmology Cambridge, Madingley Road, Cambridge, CB3 0HA, U.K.\\}

\date{Accepted ---. Received ---; in original form \today}
\pagerange{\pageref{firstpage}--\pageref{lastpage}}
\pubyear{2011}

\begin{document}

\maketitle

\begin{abstract}
Powellsnakes \citep[PwS;][]{PwSI} is a Bayesian algorithm for
detecting compact objects embedded in a diffuse background, and was
selected and successfully employed by the Planck consortium
\citep[][a]{PlnkMission} in the production of its first public
deliverable: the Early Release Compact Source Catalogue
\citep[][b]{PlnkERCSC} (ERCSC).  We present the critical foundations
and main directions of further development of PwS, which extend it in
terms of formal correctness and the optimal use of all the available
information in a consistent unified framework, where no distinction is
made between point sources (unresolved objects), SZ clusters, single
or multi-channel detection.  An emphasis is placed on the necessity of
a multi-frequency, multi-model detection algorithm in order to achieve
optimality.
%{\color{red} and include details of the next code version to be
%released, code version 4.}
\end{abstract}
\begin{keywords}
Cosmology: observations -- methods: data analysis -- cosmic microwave
background
\end{keywords}

\section{Introduction}
\label{sect:PwSIIIntro}

The detection and characterisation of discrete objects is a common
problem in many areas of astrophysics and cosmology.  Indeed, every
data reduction process must resort to some form of compact object
detection, since either the objects themselves are the goal of the
study or they act as contaminants and therefore must be removed.  In
such analyses, the key step usually involves the separation of a
localised object signal from a diffuse background, defined as all
contributions to the image aside from the objects of interest.

A well-established method to address this issue is to
  assume that most of the pixels are part of the background
  exclusively
\footnote{This is possible only if the fields are not very densely
  packed with objects.}, the background is smoothly varying, i.e. has
a characteristic length-scale much larger than that of the objects of
interest and the object are bright compared with the background.
A successful example of an object detection algorithm
based on these assumptions is SExtractor \citep{SexT}.  Its first step
is to estimate the image background. The algorithm builds up an
intensity histogram iteratively and clips it around its median.  The
resulting mesh (resembling a `swiss-cheese') is then bilinearly
interpolated to fill in the holes.  After this background has been
subtracted, the detection and characterisation of the objects is
performed either by looking for sets of connected pixels above a given
threshold or by boosting the image maxima with the help of an
`on-the-fly' convolution using a pre-defined kernel or the beam PSF.

Despite their general acceptance, such methods run into
difficulties when the characteristic extension of the fluctuations of the
diffuse background match the size and the amplitudes of the
objects of interest. Moreover, problems also arise when dealing with
low or very low signal-to-noise-ratio (SNR) data, when the rms level
of the background is comparable to, or even somewhat larger than, the
amplitude of the localised objects of interest.  A good example of
this situation is the detection of the Sunyaev-Zel'dovich
\citep{SZEfectOrg} (SZ) effect in galaxy clusters, which have
characteristic scales similar to that of the primordial CMB emission,
and at the same time are very faint and extended.  In such cases,
traditional methods fail to provide a statistically supported
prediction about the uncertainties on the parameter estimates.

The standard approach for dealing with such difficulties is to employ
linear filtering, which is an extremely well-developed field, very
firmly rooted in the principles of the orthodox school of statistics
and signal processing \citep{VTrees}.  These methods usually start by
applying a linear filter $\psi(\vect{x})$ to the original image
$d(\vect{x})$, and instead analyse the resulting filtered field.  The
filter is most often constructed by assuming a given (possibly
parametrised) spatial template, $\tau(\vect{x})$, for the objects of
interest. Depending on the application, this profile may contain
parameters (to be estimated) and already include the beam spreading
effects. The common design goals for the filter follow the
traditional, orthodox figures of merit: unbiasedness and efficiency.
The optimal solution under these constraints is well-known to be the
matched filter \citep[MF;][]{MFilter}. One may consider the filtering
process as optimally boosting (in a linear sense) the signal from
discrete objects, while simultaneously suppressing the emission from
the background. The filtering methodology has yet another major
advantage of being extremely fast and very simple to implement using
`off-the-shelf' routines (such as FFTs).  The uncertainties in the
parameter estimates are usually obtained from simulations.  In
practice, however, implementations of the filtering codes must be
supported by ancillary steps in order to cope with the artifacts
introduced as a consequence of the statistical description of the
detection process being incomplete \citep{CaniegoRings,MelinPmf}.

A natural evolution of the MF, the matched multi-filter (MMF), follows
exactly the same underlying principles and extends them to
multi-channel data sets \citep{HerranzMMF,LanzMMF}.  The simultaneous
multi-frequency analysis of a set of images has the immediate
advantage of exploiting the objects' distinctive spectral signature,
if any. Two further advantages of this technique are: (i) it boosts
the signal from the objects of interest simply by adding more data;
and (ii) it improves the elimination of the background components by
taking advantage of their correlation between channels. Once again,
the thermal SZ effect embedded in primordial CMB emission provides a
very good example.  Owing to the well-defined and unique frequency
dependence of the SZ effect, it is possible to design a filter that
combines multi-frequency maps to make possible the extraction of deep
catalogues even if the SZ component is sub-dominant in all the
channels \citep[][c]{PlnkEarlySZ}.

Further development of traditional filtering techniques includes the
`scale-adaptive filter' \citep[SAF;][]{SanzSAF,HerranzSAF}, in which
the physical scale of the objects of interest is added as an extra
degree of freedom and an additional condition for optimality is added
in the derivation of the filter. \cite{SchafferII} generalised the
scale-adaptive filter to the spherical topologies and added
multi-channel support.

A very popular member of the filter family is the wavelets group, in
particular the mexican-hat (MexHat) wavelet of order $1$. Indeed, this
MexHat wavelet is the MF or the SAF solution under particular
assumptions about the statistical properties of the background and the
objects profile \citep{SanzSAF}.  Since such conditions hold very well
in modern cosmological data sets, such as those obtained from WMAP
\citep{WMAPI} or Planck, and the simplicity of the function allows
easy and robust engineering, the MexHat wavelet has been a favourite
detection tool of many authors \citep{MexHat1,MexHatII}. Nonetheless,
obtaining good results with the MexHat filter is extremely dependent
on the value of the acceptance/rejection threshold.  The only way to
ensure optimal performance is to run the code on realistic simulations
and then assess the code's yield against the simulation's input
catalogue, but a large number of runs is needed to fine-tune the
threshold value.  Exactly the same procedure must be followed to
determine the uncertainties on the parameter estimates. This may not
seem a severe limitation, since immense computing resources are now
cheaply available. Given the increased level of accuracy and
complexity of current cosmological data sets, however, simulations
must be rather sophisticated to provide a realistic test bed, and so
even the enormous computational resources available are not sufficient
to cope with the massive throughput demanded. For example, a single
realistic Planck simulation (FFP) takes about one full week to run on
a very large cluster and to have reasonable estimates of the parameter
uncertainties and detection thresholds, at least several hundred
independent simulations are needed.

To overcome these limitations of linear filtering methods,
\cite{mikecharlie} introduced a detection algorithm based on a
Bayesian approach. As with the filtering techniques, the method
assumed a parameterised form for the objects of interest, but the
optimal values of these parameters, and their associated
uncertainties, were obtained in a single step by evaluating their full
posterior distribution. Another major advantage of this method is the
consistent inclusion of physical priors on the parameters defining the
objects and on the number of objects present, which improve the
detection efficiency. Although this approach represented a further step
in the direction of bringing a more solid statistical foundation to
the object detection/characterization problem, its implementation was
conducted using a Monte-Carlo Markov chain (MCMC) algorithm to sample
from a very complex posterior distribution with variable
dimensionality (dependent on the number of objects). This technique
therefore proved extremely computational intensive. Despite the
considerable progress that has recently been made towards increasing
the efficiency of sampling-based Bayesian object detection methods
\citep{HobsonMCMC2007}, such algorithms are still substantially slower
than simple linear filtering methods. In a recent work, \cite{Argueso}
suggested a semi-analytical hybrid Bayesian {\em maximum a posteriori}
(MAP) scheme to overcome the complexity and the massive resources
required the Hobson \& McLachlan method. Its main advantage is really
simplicity. However, the method still relies on the MF to find the
source's positions and, therefore, it does not embody the full
Bayesian logic. Meanwhile, \cite{PwSI} and \cite{MultiNest} have moved
one step further towards the theoretically-optimal Bayesian solution
by exploring the use of evidence ratio methods, which are the optimal
decision-making tools (see section \ref{sect:DecisionTheory}),
rather than simply adopting the MAP solution.

Our proposal here is to blend detection strategies, i.e.
multi-channel filtering, Bayesian posterior sampling and evidence
ratio evaluation, into a rigorous, hybrid, multi-model scheme (as
opposed to traditional binary models). This novel methodology is
simultaneously general, formally and statistically firmly grounded,
and overcomes the computation inefficiencies of the pure sampling
methodologies. This is PowellSnakes II.

The structure of this paper is as follows. In
Section~\ref{sect:BayInf}, we give an overview of Laplace--Bayes
probability theory and its close relationship with decision theory as
a consistent inference and decision-making device. Our data
model and the different constituents of the Bayesian framework, namely
the likelihood and priors are discussed in
Section~\ref{sect:BayObjectDetection}, and in
Section~\ref{sect:ObjectDetectStrat} we bring together these elements
and recommend an implementation strategy based on the exploration of
the properties and symmetries of the posterior manifold. We also
identify problems that may arise and we suggest
effective ways of tackling them using the Bayesian formalism.
Finally we present our conclusions and directions for future work in
Section~\ref{sect:Conclusions}.

%***-------------------------------------------------------------***

\section{Bayesian inference}
\label{sect:BayInf}
\subsection{Basic tools}
\label{sect:BayInfBasicTool}
The Bayesian system of inference is the only one that provides a
consistent extension of deductive logic ( \{ 0=false, 1= true\} ) to a broader class of `\emph{degrees-of-belief}' by mapping them into the real interval $[0,1]$
\citep{jaynes}. Combining the multiplication rule together with
the associativity and commutativity properties of the logical product,
one may write the equation which will give us the posterior
probability of a set of parameters ($\vect{\Theta}$) taking into
account the data ($\vect{d}$) and the underlining hypothesis ($H$).
This equation is also known as Bayes theorem
\begin{equation}
\Pr(\vect{\Theta} | \vect{d}, H) =
\frac{\Pr(\vect{d}|\,\vect{\Theta},H)\Pr(\vect{\Theta}|H)}
{\Pr(\vect{d}|H)}, \label{eq:BI_Params}
\end{equation}
where, for brevity, we denote $\Pr(\vect{\Theta} | \vect{d}, H) \equiv
P(\vect{\Theta})$ as the posterior probability distribution of the
parameters, $\Pr(\vect{d}|\,\mathbf{\Theta},H) \equiv
\mathcal{L}(\vect{\Theta})$ as the Likelihood, $\Pr(\vect{\Theta}|H)
\equiv \pi(\vect{\Theta})$ as the prior and $\Pr(\vect{d}|H) \equiv
\mathcal{Z}$ as the Bayesian evidence. The (unnormalised) posterior
distribution is the complete inference of the parameter values
$\vect{\Theta}$, and thus plays the central role in Bayesian parameter
estimation.

The normalised posterior distribution may be easily obtained by
integrating over all possible values of the parameters and equating
the resulting expression to one, and from this argument one can easily
see that the evidence is given by
\begin{equation}
\mathcal{Z} \equiv \Pr(\vect{d} | H) =
\int{\mathcal{L}(\vect{\Theta})\,\pi(\vect{\Theta})}\,d^K\vect{\Theta},
 \label{eq:BI_EvidDef}
\end{equation}
where $K$ is the dimensionality of the parameter space. Inspecting
this expression, one immediately recognizes that the evidence is the
expectation of the likelihood over the prior, and hence is central to
Bayesian model selection between different hypothesis $H_i$.  We note
that the evidence evaluation requires the prior to be properly
normalised.

\subsection{Decision theory}
\label{sect:DecisionTheory}

Probability theory defines only a state of knowledge: the posterior
probabilities. There is nothing in probability theory \textit{per se}
that determines how to make decisions based on these probabilities.
Indeed, a range of actions is always possible, even when using the
same state of knowledge, because the cost of making a wrong decision
usually changes according to the kind of problem under analysis.  For
example, in the case of object detection, one often considers each
type of error, i.e. an undetected object or a spurious detection, as
equally bad. For a moment, however, suppose we instead wished to
determine whether or not a certain person was immune to a certain
pathogen. Failing to detect a previously acquired immunity would only
cost the price of an extra vaccine, but failing to determine that
someone was not immune could seriously put her/his life at risk. Thus,
even with the same degree of knowledge, the cost of choosing
incorrectly is not the same in every case. To deal with such
difficulties, one must apply decision theory (DT), which we now
briefly summarise.

To apply decision theory, one must first define the {\em loss/cost
  function} $L(\vect{D},\vect{E})$ for the problem at hand, where
$\vect{D}$ is the set of possible decisions and $\vect{E}$ is the set
of true values of the entities one is attempting to infer. In general,
these entities can be either continuous parameters or discrete
hypotheses, and so decision theory can be applied equally well both to
parameter estimation and model selection.  The loss function simply
maps the `mistakes' in our estimations/selections, $\vect{D}$, into
positive real values $L(\vect{D},\vect{E})$, thereby defining the
penalty one incurs when making wrong judgments. The Bayesian approach
to decision theory is simply to minimise, with respect to $\vect{D}$,
the expected loss
\begin{equation}
\langle L(\vect{D},\vect{E})
\rangle = \int\!\!\!\int
L(\vect{D},\vect{E})\, \Pr(\vect{D},\vect{E})\,d\vect{D}\,d\vect{E}.
\label{eq: BayInf_AveLoss}
\end{equation}

\subsubsection{Parameter estimation}
\label{sect:ParamEstimation}

In the estimation of a set of continuous parameters\footnote{
In astronomical object detection, the majority of the interesting
parameters (flux, position, geometry, etc.) are continuous and real
valued. Moreover, discrete parameters can always be handled within the
same framework by resorting to delta Dirac functions.}, the `decisions'
$\vect{D}$ are the parameter estimates $\widehat{\vect{\Theta}}$ and the
`entities' $\vect{E}$ are the true values $\vect{\Theta}^\ast$ of the
parameters. Typically, the loss function is taken to be a
function of the difference, or error, $\epsilon \equiv
\widehat{\vect{\Theta}} - \vect{\Theta}^\ast$.

Some popular choices of loss functions are: (i) the square error
$\epsilon^2$; (ii) the absolute error $|\epsilon|$; and (iii) the
uniform cost inside error bar, i.e. unity if $|\epsilon| > \Delta$ and
zero if $|\epsilon| < \Delta$, where $\Delta$ is some pre-defined
small quantity. In each case, one can easily find the optimal
estimator by minimising the expected loss (\ref{eq: BayInf_AveLoss})
with respect to $\widehat{\vect{\Theta}}$. The solutions are,
respectively: (i) the posterior mean; (ii) the posterior median; and
(iii) the posterior mode.\footnote{The posterior mean is the Bayesian
  optimal estimator under a very broad class of reasonable loss
  functions. When the posterior distribution is Gaussian all three
  common estimators match, and the posterior mode is often the
  simplest to compute. Nonetheless, if the parameter space is, in
  practice, discrete (e.g. pixelisation), the posterior mean might
  provide a hyper-resolution estimate (sub-pixel accuracy).}

%It is worth noting that, in contrast, the frequentist approach to
%statistics does not have any consistent and well-defined criterion for
%defining the optimality of an estimator. Instead, estimators are
%ranked in quality according to \emph{ad hoc} criteria such as bias and
%efficiency. Indeed, it is well known that some unbiased estimators are
%sub-optimal in the decision-theory sense \citep{jaynes}.

The most popular choice of loss function among the astronomical
community is the square error $\epsilon^2$. When detecting
astronomical objects, however, the requirements are usually not those
of the square error function, which puts an extreme emphasis on values
very far from the true ones. This extreme sensitivity to the outliers
makes the posterior mean estimator less robust than, for example, the
posterior median, which is much more resilient to outliers. An even
better choice would be not penalise the estimates at all if they fall
within a small neighbourhood $\Delta$ around the true parameters
values and prescribe a constant penalty otherwise.  This is precisely
the `uniform cost inside error bar' loss function decribed above.
This loss criterion closely matches what we would intuitively expect
when assessing the quality of a detection algorithm. For example, if
the estimated value of a source flux lies outside the allowed range it
does not matter how far it lies from the true value, since it will
always be counted as a spurious detection \citep[][b]{PlnkERCSC}.

\subsubsection{Interval estimation}

In addition to an estimate $\hat{\vect{\Theta}}$, one typically
summarises the inference implied by the full posterior distribution by
quoting either joint or marginalised confidence intervals (or, more
precisely, Bayesian credible intervals). One could, in principle,
obtain an optimal interval by employing an approprite loss function,
but a simpler approach is now widely accepted, namely the high
probability density interval (HPD). The HPD interval containing the
fraction $(1 - \alpha)$ of the total probability is defined such that:
$\Pr(\vect{\Theta} \in \mbox{HPD}|\vect{d},H) = 1-\alpha$ and, if
$\vect{\Theta}_1 \in \mbox{HPD}$ and $\vect{\Theta}_2 \, \not\in
\mbox{HPD}$, then $\Pr(\vect{\Theta}_1 |\vect{d},H) \geq
\Pr(\vect{\Theta}_2 | \vect{d},H)$.

The characterization of the HPD interval may be easily obtained by
sampling from the posterior distribution.  When the posterior
distribution is known to be Gaussian or close to it, which is a very
common case, the $\pm$rms interval is usually quoted instead.

\subsubsection{Model selection and catalogue making}
\label{sect:DecisionTheoryCatMaking}

In model selection, the decision theory `entities' $\vect{E}$ are the
hypotheses under consideration and the `decisions' $\vect{D}$ are the
chosen hypotheses, such that $L(D_i,H_j) \equiv L_{ij}$ is the loss
associated with the decision $D_i \equiv \mbox{choose $H_i$}$, when
$H_j$ is true. Thus, inserting this form for the loss matrix into the
right hand side of equation (\ref{eq: BayInf_AveLoss}) and performing
the integration using the delta Dirac functions to represent
discrete values as infinite densities, the average loss reads
\begin{equation}
\langle
L(\vect{D},\vect{H})
\rangle  = \sum_{ij} L_{ij}\, \Pr(D_i, H_j). \label{eq:AveLossDisc}
\end{equation}

If, for example, one is interested in distinguishing between a null
hypothesis $H_0$ and a given alternative hypothesis $H_1$ from amongst
a wider collection, then typically the loss function has the form
\begin{equation}
\label{eq:LossMatrixHyp1}
L_{ij} =
\left\{
\begin{array} {ll}
0 & \mbox{if $i=j$ (no penalty if correct)}\\
\mbox{positive value} & \mbox{if $i=1$, $j\neq i$ (false positive)} \\
\mbox{positive value} & \mbox{if $j=1$, $i \neq j$ (false negative)}\\
0 & \mbox{otherwise (alternative selection error)}
\end{array}
\right.
\end{equation}
Minimizing (\ref{eq:AveLossDisc}) is not a difficult task
\citep{VTrees}, but the general case above leads to long and
cumbersome expressions that we shall not explore now.

Much simpler and enlightening, but still capable of a very broad and
interesting range of practical applications, is the binary case
consisting of just two hypotheses $H_0$ and $H_1$. In this case, the
decision criterion that minimises the expected loss is
\begin{equation}
\label{eq:AveLossCatMak1}
    \ln \left[ \frac{\Pr(H_1 | \vect{d})}{\Pr(H_0 | \vect{d})} \right] \overset{H_1}{\underset{H_0}{\gtrless}} \xi
\end{equation}
where $\xi \equiv \ln \frac{L_{10}}{L_{01}}$.
The `posterior odds' ratio
\begin{equation}
\frac{\Pr(H_1 | \vect{d})}{\Pr(H_0 | \vect{d})} =
\frac{\mathcal{Z}_1}{\mathcal{Z}_0} \, \frac{\Pr(H_1)}{\Pr(H_0)}
\label{eq:podds}
\end{equation}
gives the posterior probabilities of the models given the data and is
a very commonly-used quantity in Bayesian model selection.  Indeed,
Jaynes asserts that the best way to decide between two models is by
computing the posterior odds and compare it against a threshold.
Using DT we have recovered this result and, at the same time, given it
a precise statistical meaning and also defined a threshold for
decision making based on the loss criterion.

Unfortunately, in astronomy it is often not possible to assign
meaningful values to the loss. In particular, in object detection and
catalogue making, astronomers like instead to measure the quality of a
catalogue in terms of the expected/maximum contamination (false
positive rate) and the expected/minimum completeness (true positive
rate). There is, of course, a connection between this approach and DT,
but quantifying it is not trivial. Nonetheless, there a very simple
and powerful way to to define the acceptance/rejection threshold in
Bayesian catalogue making, based on the probabilities of the different
errors that might occur (i.e. spurious or missed detections), but we
shall postpone its discussion until
Section~\ref{sect:ObjectDetectStrat}.

Before moving on, it is worth mentioning that, if one ignores the
(often crucially important; \citep[][ch. 30, pag. 1132]{MikesBook})
 factor $\Pr(H_1)/\Pr(H_0)$ in (\ref{eq:podds}),
the remaining evidence ratio $\mathcal{Z}_1/\mathcal{Z}_0$ depends only on the data and can thus be
viewed as an orthodox statistic. As such, the properties of its
sampling distributon can be investigated using standard frequentist
tools, such as the `power' (true positive rate) $\Pr(D_1|H_1)$ and the
`type I error rate' (false positive rate) $\Pr(D_1|H_0)$
\citep{PeacockFriend}. Such analyses overlook, however, that the
evidence ratio {\em is} the optimal decision rule. The only degree of
freedom remaining is the choice of a threshold, which determines
whether it is preferable to have fewer (more) detections at the cost
of good (poor) rejection; there is no way of decreasing both error
rates simultaneously because the evidence ratio is already the most
discriminating statistic. The claim by \cite{PeacockFriend} that the
evidence ratio test is not `powerful' results from them fixing the
threshold in an arbitrary way; it is this threshold that controls the
balance between different error rates, and not the statistic itself.
A better way of measuring the quality of a binary classifier based on
some statistic is to allow the threshold to vary and plot the
resulting true positive rate against the false positive rate. This
produces the receiver operating characterisic (ROC) curve of the
classifier. A common measure of classifer quality is the area under
the ROC-curve (the AUC statistic), which obviously does not rely on
choosing a single threshold. One may show that the AUC is equal to the
probability that the classifier will rank a randomly chosen data set
generated from $H_1$ higher than a randomly chosen data set generated
from $H_0$.

\section{Bayesian object detection}
\label{sect:BayObjectDetection}

\subsection{Data model}
\label{sect:BayObjectDetectionStatModel}

The specification of the PwS statistical model for a single-frequency
observation of localised objects embedded in a background is given in
\cite{PwSI}. This can be straightforwardly extended to accommodate
multi-frequency observations. At each observing frequency $\nu$, PwS
treats the observed data $d_\nu(\vect{x})$ as the superposition of a
`generalised' noise background $n'_\nu(\vect{x}) = b_\nu(\vect{x}) +
n_\nu(\vect{x})$, consisting of background sky emission and
instrumental noise, plus a characteristic signal $s_\nu(\vect{x})$
coming from the sources. For ease of notation, we will collect the
fields at different frequencies into vectors. Moreover, the signal and
background components in each frequency channel are assumed to have
been smoothed with a known beam, which may differ between
channels. The resulting model for the data vector $\vect{d}(\vect{x})$
reads
\begin{equation}
\label{ed:SourcesModel0}
\vect{d}(\vect{x}) = \sum^{N_{{\mathrm{s}}}}_{j=1}
\vect{s}_j(\vect{x};\vect{\Theta}_j) + \vect{b}(\vect{x}) +
\vect{n}(\vect{x}),
\end{equation}
where $N_{\mathrm{s}}$ is the number of sources,
$\vect{s}_j(\vect{x};\vect{\Theta}_j)$ is the signal vector due to the
$j$th source, which depends on the parameter vector $\vect{\Theta}_j$
characterising the object, $\vect{b}(\vect{x})$ is the signal vector
due to the diffuse astronomical backgrounds, and $\vect{n}(\vect{x})$
is the instrumental noise vector. The astronomical backgrounds denoted
by $\vect{b}(\vect{x})$ are expected to exhibit strong correlations
between different frequency channels, whereas the instrumental noise
$\vect{n}(\vect{x})$ is expected to be uncorrelated between frequency
channels, and also between pixels in the case of simple white
noise.\footnote{The condition of the instrumental noise being white is
  not necessary.  The general case of correlated noise between pixels
  does not complicate the mathematical development, but can increase
  computational expense. In any case, the assumption of white noise
  applies extremely well to Planck data on the small scales of
  interest used for the identification of localised objects.}

We write the signal vector due to the $j$th source in
(\ref{ed:SourcesModel0}) as
\begin{equation}
\label{ed:SourcesModel1}
\vect{s}_j(\vect{x};\vect{\Theta}_j) = A_j \vect{f}(\vect{\phi}_j)
\vect{\tau}(\vect{x}-\vect{X}_j;\vect{a}_j),
\end{equation}
where the vector $\tau(\vect{x}-\vect{X}_j;\vect{a}_j)$ denotes the
convolved spatial template at each frequency of a source centred at
the position $\vect{X}_j$ and characterised by the shape parameter
vector $\vect{a}_j$, the vector $\vect{f}$ contains the emission
coefficients at each frequency, which depend on the emission law
parameter vector $\vect{\phi}_j$ of the source (see below), and $A_j$
is an overall amplitude for the source at some chosen reference
frequency. Thus, the parameters to be determined for the $j$th source
are its overall amplitude, position, shape parameters and emission law
parameters, which we denote collectively by $\vect{\Theta}_j =
\{A_j,\vect{X}_j,\vect{a}_j,\vect{\phi}_j\}$. The totality of these
parameters, for all the sources present, plus the number of sources
$N_{\mathrm{s}}$, are concatenated into the
single parameter vector $\vect{\Theta}$.  For convenience, we denote
the signal vector generated by all the sources by
\begin{equation}
\label{eq:SourceDef0}
\vect{s}(\vect{x};\vect{\Theta}) \equiv
\sum^{N_{\mathrm{s}}}_{j=1} \vect{s}_j(\vect{x};\vect{\Theta}_j).
\end{equation}

The nature of the emission law parameter vector $\vect{\phi}$ depends
on the class of object under consideration. PwS analyses the data
assuming that all the objects belong to a single class, and repeats
the analysis for each class of interest. The assignment of individual
sources to a particular class is then performed via a model selection
step (see Section~\ref{sect:ObjStratMultiModel}). The number and specification of classes can
be arbitrary, including, for example, SZ clusters, point sources,
Galactic objects, etc.\ Previous multi-frequency versions of PwS have
been limited to the case where all objects share the same, fixed
emission law. SZ clusters fall exactly in this category as, ignoring
the relativistic corrections, they all follow exactly the same
spectral signature \citep{SZBirkinshaw}, which does not depend on any
parameters. For extragalactic point sources, however, the emission law
is phenomenological and can vary between sources. Consequently,
PwSII has been extended to accommodate such cases. For example, two
important families of extragalactic point sources in Planck data are
as follows.
\begin{itemize}
  \item Radio sources are the dominant family of point sources for all
    Planck channels up to and including 143 GHz.  Based on the work of
    \cite{WaldramPS}, we assume an emission law for such objects of
    the form
  \begin{equation}
  \label{eq:PriorsSpectralRadio0}
    \ln f_{\nu} = \alpha ~ \ln \left(\frac{\nu}{\nu_0} \right) + \beta ~ \left[\ln \left(\frac{\nu}{\nu_0} \right)\right]^2,
  \end{equation}
where $\vect{\phi} = \{\alpha,\beta\}$ are spectral parameters
  that can vary between sources, and $\nu_0$ is the reference
  frequency (note that $f_\nu = 1$ at $\nu=\nu_0$). Setting $\beta=0$,
  recovers the commonly-assumed power-law spectral behaviour with
  spectral index $\alpha$. The more general form
  (\ref{eq:PriorsSpectralRadio0}) accommodates most of the common types
  of radio-source spectra, namely: flat, steep, and inverted.
  \item Dusty galaxies dominate
    the Planck highest frequency channels, starting at 217 GHz up to
    857 GHz. Their spectral behaviour may be represented to very
    good accuracy using the well-known greybody  model
  \begin{equation}
  \label{eq:PriorsSpectralDusty0}
    \ln f_{\nu} =  \beta ~ \ln \left(\frac{\nu}{\nu_0} \right)
+ \ln \left[\frac{B_\nu(T)}{B_{\nu_0}(T)}\right],
  \end{equation}
  where the spectral parameters $\vect{\phi}=\{\beta,T\}$ are the dust
  emissivity and temperature respectively, $B_\nu(T)$ is the Planck law of blackbody radiation and
  $\nu_0$ is once again the reference frequency  \citep{GreyBody}. We have again
  normalised (\ref{eq:PriorsSpectralDusty0}) such that $f_\nu=1$ at
  $\nu=\nu_0$.
\end{itemize}

\subsection{Likelihood}
\label{sect:BayObjectDetectionLikelihood}

The form of the likelihood is determined by the statistical properties
of the generalised noise (background sky emission plus instrumental
noise) in each frequency channel. As in PwSI, we will perform our
analysis in sufficiently small patches of sky such that it is not
unreasonable to assume statistical homogeneity. In this case, it is
more convenient to work in Fourier space, since there are no
correlations between the Fourier modes of the generalised noise, which
leads to considerable savings in computation and storage.  Moreover,
we will assume that both the background emission and instrumental
noise are Gaussian random fields. This is a very accurate assumption
for instrumental noise and the primordial CMB, but more questionable
for Galactic emission.

We are, in fact, interested only in the likelihood ratio between the
hypothesis $H_\mathrm{s}$ that objects (of a given source type $s$)
are present and the null hypothesis $H_0$ that there are no such
objects. The latter corresponds to setting the sources signal
$\vect{s}(\vect{x};\vect{\Theta})$ to zero. Under our combined
assumptions, the log-likelihood ratio has the form
\begin{eqnarray}
\label{eq:LikeRatioFinal}
\ln\left[\frac{\mathcal{L}_{H_\mathrm{s}}(\vect{\Theta})}{\mathcal{L}_{H_0}(\vect{\Theta})}\right]
& = & \sum_{\vect{\eta}}
\widetilde{\vect{d}}^t(\vect{\eta})\vect{\mathcal{N}}^{-1}(\vect{\eta})
\widetilde{\vect{s}}(\vect{\eta};\vect{\Theta}) \nonumber \\ && -
\tfrac{1}{2} \sum_{\vect{\eta}}
\widetilde{\vect{s}}^t(\vect{\eta};\vect{\Theta})
\vect{\mathcal{N}}^{-1}(\vect{\eta})\widetilde{\vect{s}}(\vect{\eta};\vect{\Theta}),
\label{eq:LikeRatioFinal}
\end{eqnarray}
where the tilde denotes a Fourier transform, the usual mode wavenumber
$\vect{k}=2 \pi\vect{\eta}$, and the matrix
$\vect{\mathcal{N}}(\vect{\eta})$ contains the generalised noise
cross-power-spectra.

From (\ref{ed:SourcesModel1}) and (\ref{eq:SourceDef0}), the Fourier transform of the signal due to all the
sources may be written
\begin{equation}
\widetilde{\vect{s}}(\vect{\eta};\vect{\Theta}) =
\widetilde{\vect{B}}(\vect{\eta})\sum_{j=1}^{N_{\mathrm{s}}} A_j \vect{f}(\vect{\phi}_j)
\widetilde{\tau}(-\vect{\eta};\vect{a}_j)\textrm{e}^{\textrm{i} 2\pi{\bm \eta}\cdot\vect{X}_j},
\label{eq:sft}
\end{equation}
where the vector $\widetilde{\vect{B}}(\vect{\eta})$ contains the
Fourier transform of the beam at each frequency and
$\widetilde{\tau}(\vect{\eta};\vect{a})$ is the Fourier transform of
the template for an unconvolved object at the origin, characterised by
the shape parameters $\vect{a}$.

Substituting (\ref{eq:sft}) into (\ref{eq:LikeRatioFinal}) and
rearranging, one obtains the final form for
the likelihood ratio, which we will use throughout, namely
\[
\hspace*{-6cm}\ln\left[\frac{\mathcal{L}_{H_\mathrm{s}}(\vect{\Theta})}{\mathcal{L}_{H_0}(\vect{\Theta})}\right] =
\]
\vspace*{-0.4cm}
\[
\sum^{N_{\mathrm{s}}}_j \left\{A_j\mathcal{F}^{-1}
\left[\mathcal{P}_j(\vect{\eta})
\widetilde{\tau}(-\vect{\eta};\vect{a}_j)\right]_{\vect{X}_j} - \tfrac{1}{2}A_j^2 \sum_{\vect{\eta}}
\mathcal{Q}_{jj}(\vect{\eta})
|\widetilde{\tau}(\vect{\eta};\vect{a}_j)|^2\right\}
\]
\vspace*{-0.4cm}
\begin{equation}
- \sum^{N_{\mathrm{s}}}_{i > j}\left\{A_iA_j
\mathcal{F}^{-1}
\left[\mathcal{Q}_{ij}(\vect{\eta})
\widetilde{\tau}(\vect{\eta};\vect{a}_i)
\widetilde{\tau}(-\vect{\eta};\vect{a}_j)\right]_{\vect{X}_i
- \vect{X}_j} \right\},
\label{eq:LikeFilterComplete}
\end{equation}
where $\mathcal{F}^{-1}[\ldots]_\vect{x}$ denotes the inverse Fourier
transform of the quantity in brackets, evaluated at the point
$\vect{x}$, and we have defined the quantities $\mathcal{P}_j({\bm
  \eta}) \equiv \widetilde{\vect{d}}^t(\vect{\eta})
\vect{\mathcal{N}}^{-1}(\vect{\eta}) \vect{\psi}(\vect{\eta})$ and
$\mathcal{Q}_{ij}({\bm \eta}) \equiv
\widetilde{\vect{\psi}}_i^t(\vect{\eta})
\vect{\mathcal{N}}^{-1}(\vect{\eta}) \vect{\psi}_j(\vect{\eta})$, in
which the vector $\psi_i(\vect{\eta})$ has the components
$(\vect{\psi}_{i})_\nu = \widetilde{B}_\nu(\vect{\eta})
(\vect{f}_{i})_\nu$, with $\nu$ labelling frequency channels.

We have written the likelihood ratio in this way since it combines
multi-channel data into a single equivalent channel. Moreover, it
highlights the importance of the final `cross-term' on the RHS of
(\ref{eq:LikeFilterComplete}). Let us assume for a moment that this
cross-term is negligible. In this case, the parameters of each source
enter the likelihood independently. This parameter independence allows
us to perform our analysis one source at a time and forms the basis of
the `single source model' discussed in Section~\ref{sect:ObjStratDetection}, which greatly
simplifies the source detection problem. The physical meaning of the
neglected cross-term is most easily understood by considering the
simple, but important, example of point sources, for which
$\tau(\vect{x},\vect{a}) = \delta(\vect{x})$. In this case, the
cross-term in (\ref{eq:LikeFilterComplete}) becomes
\begin{equation}
\sum_{i>j} A_i A_j
\mathcal{F}^{-1}[\mathcal{Q}_{ij}(\vect{\eta})]_{\vect{X}_i-\vect{X}_j}.
\label{eq:WellSeparetedObjects}
\end{equation}
A sufficient condition for this expression be small is that all the
sources are sufficiently well-separated that $\mathcal{F}^{-1}[\mathcal{Q}_{ij}(\vect{\eta})]_\vect{x}$ is close to zero for such distances.
For simple, uncorrelated backgrounds, $\mathcal{Q}_{ij}(\vect{\eta})$
contains just linear combinations of the instrument beams in each
frequency channel. Thus, the condition that
(\ref{eq:WellSeparetedObjects}) is small is just a generalization of
the common assumption in astronomy that objects are well separated, or
that object blending effects are negligible
\footnote{When the background is uncorrelated, this condition is
  immediately fulfilled if each pixel contains signal coming from one
  and only one source. However, this is not the case when there are
  strong correlations in the background as in the case of Planck.}.
When detecting point sources, and assuming the blending is not severe,
an efficient implementation of the full deblending term is possible,
but this will be addressed in a forthcoming publication.

It is worth noting that maximising the likelihood ratio
(\ref{eq:LikeFilterComplete}), in the absence of the cross-term
(\ref{eq:WellSeparetedObjects}), with respect to the source amplitudes
$A_j$, gives
\begin{equation}
\label{eq:MaxLikeEstimate1}
\widehat{A}_j = \frac{\mathcal{F}^{-1} \left[\mathcal{P}_j(\vect{\eta})
\widetilde{\tau}(-\vect{\eta};\widehat{\vect{a}}_k)\right]_
{\widehat{\vect{X}}_j}}
{\sum_{\vect{\eta}} \mathcal{Q}_{jj}(\vect{\eta})
|\widetilde{\tau}(\vect{\eta};\widehat{\vect{a}}_j)|^2},
\end{equation}
which recovers the expression for the matched multi-filter (MMF)
\citep{HerranzMMF}. Thus, we see that the filtered field is merely the
projection of the likelihood manifold onto the subspace of position
parameters $\vect{X}_j$. This identification further allows one
straightforwardly to estimate the uncertainties on all the MMF
parameter estimates simultaneously by calculating and inverting the
Hessian matrix of the likelihood at its peak(s). This should be
contrasted with traditional approaches to MMF in which the uncertainty
on the estmated source flux is calculated assuming the values of all
other parameters are fixed \citep{MelinPmf}.

Moreover, substituting the maximum-likelihood estimate
(\ref{eq:MaxLikeEstimate1}) into the expression
(\ref{eq:LikeFilterComplete}) for the likelihood ratio, one obtains
for the $j$th object
\begin{equation}
\label{eq:MaxLikeEstimate2}
\mathrm{max}\left[\ln\left(\frac{{\mathcal{L}_{H_\mathrm{s}}}}{\mathcal{L}_{H_0}}\right) \right] =
\tfrac{1}{2} \sum_{{\bm \eta}} \mathcal{Q}_{jj}(\vect{\eta})
|\widetilde{\tau}(\vect{\eta};\widehat{\vect{a}}_j)|^2 \widehat{A}_j^2
=
\tfrac{1}{2}\widehat{\mathrm{SNR}}_j^{~2}
\end{equation}
where $\widehat{\mathrm{SNR}}_j$ is the signal-to-noise ratio (at the
peak) of the $j$th source, and the rms $\sigma$ of the noise satisfies
\begin{equation}
\label{eq:MaxLikeEstimateSigma}
\frac{1}{\sigma^2} = \sum_{\vect{\eta}} \mathcal{Q}_{jj}(\vect{\eta})
|\widetilde{\tau}(\vect{\eta};\widehat{\vect{a}}_j)|^2.
\end{equation}
Thus, one sees that in the traditional approach to catalogue making,
in which one compares the maximum SNR of the putative detections to
some threshold, one is really performing a generalised likelihood
ratio test.

\subsection{Priors}
\label{sect:Priors}

If the data model provides a good description of the observed data and
the signal-to-noise ratio is high, then the likelihood will be very
strongly peaked around the true parameter values and the prior will
have little or no influence on the posterior distribution. At the
faint end of the source population, however, priors will inevitably
play an important role. Moreover, since for most cases in astronomy
the faint tail overwhelmingly dominates the population, the selection
of the priors becomes important and has to be addressed very
carefully.

PwSII separates the tasks of source detection (deciding whether a
certain signal is due to a source) and source estimation (determining
the parameters of the source). This separation has the advantage of
allowing the use of different sets of priors at each stage.
Typically, we first perform the source detection step using
`informative' priors, which encompass all the available information,
since they provide the optimal selection criterion and the optimal
estimators. After the set of detections has been decided, PwS proceeds
to the estimation pass, in which `non-informative' priors may be used
instead.

Non-informative priors are constructed such that the maximum a
posteriori (MAP) estimator of any quantity should depend exclusively
on the data.\footnote{These priors usually need not be properly
  normalised, since one wishes only to locate the maximum of the
  posterior distribution and the normalisation does not depend on any
  parameters.}  One way of expressing this condition is that, when
changing the data, the likelihood shape remains unchanged and only its
location in the parameter space changes \citep{BoxTiao}. Thus, the
idea is to find an appropriate re-parametrization of the likelihood
that transforms the parameters into location parameters, for which the
ignorance prior is locally uniform (locally, in this sense, means the
parameter range where the mass of the likelihood is concentrated). One
then performs the inverse parametrisation transformation on the
uniform prior to obtain the appropriate prior in the original
parameterisation.  Finding such a transformation can, however, be very
difficult for a general multi-dimensional prior.

Nonetheless, in a large majority of applications, the parameters be
may assumed independent, so that the prior factorises
\begin{equation}
\label{sect:PriorsFlux00}
\pi(\theta_1,\theta_2\ldots,\theta_n) = \pi_1(\theta_1)\pi_2(\theta_2)
\ldots \pi_n(\theta_n).
\end{equation}
For one-dimensional distributions, Jeffreys devised a general way to
derive the non-informative prior on a parameter based on invariance
properties of the likelihood under a change of variable.  The Jeffreys
rule for constructing ignorance priors for the one-dimensional case
reads
\begin{equation}
\label{eq:PriorsRadiusNI0JR}
\pi(\theta) \propto \mathcal{J}^{1/2}(\theta),
\end{equation}
where
\begin{equation}
\label{eq:PriorsRadiusNI0JR1}
\mathcal{J}(\theta) \equiv - \left\langle \frac{\partial^2 \ln
\mathcal{L}(\theta)}{\partial \theta^2}  \right\rangle
\end{equation}
is the Fisher information.  We will adopt this approach and now
consider the prior on each parameter of interest.

\subsubsection{Prior on positions}
\label{sect:PriorsPosition}

It is obvious that the distribution of sources is not uniform across
the sky. The Galactic regions (Milky Way and Magellanic Clouds) have a
much higher density of detectable sources than the rest of the sky.
Moreover, assuming extra-galactic sources to be uniformly distributed
across the sky (no clustering) is not sufficient to ensure that the
distribution of detectable sources is uniform, since the
background/noise is itself inhomogeneous over the sky.

Nonetheless, PwS divides the sky into small patches and, in each such
region, the assumptions of background homogeneity and a uniform source
distribution are reasonable. Moreover, if the sky patches used are
sufficiently small, our locally uniform model can easily cope with
clustering when the gradient of the density of sources is small across
the patch boundaries. The correctly normalised positions prior for the
complete ensemble of sources in a patch is simply
\begin{equation}
\label{eq:PriorsPosition0}
\Pr(\vect{X^{N_{\mathrm{s}}}} | N_{\mathrm{s}}, ~ N_{\mathrm{pix}}) =
\frac{1}{{N_{\mathrm{pix}}}^{N_{\mathrm{s}}}},
\end{equation}
where $N_{\mathrm{pix}}$ is the number of pixels in each patch and
$N_{\mathrm{s}}$ is the number of sources in that patch.

\subsubsection{Prior on the number of sources}
\label{sect:PriorsN}
Following the same rationale of local uniformity, i.e no clustering,
the probability of finding $N_{\mathrm{s}}$ objects (above a given
flux limit) in a sky patch follows a Poisson distribution
\begin{equation}
\label{eq:PriorsN0}
\pi(N_s) = \Pr(N_s | \lambda) = e^{-\lambda} ~ \frac{\lambda^{N_s}}{N_s!},
\end{equation}
where $\lambda$ is the expected number of such objects in that region.
Moreover, $\lambda$ should be proportional to the region size $\lambda
= \Lambda_s N_{\mathrm{pix}} \Delta_p$, where $\Lambda_s$ is the number of
sources per pixel and $\Delta_p$ is the pixel area. Note that
$\Lambda_s$ may change across the sky as we are only enforcing the
uniformity locally within each patch.

\subsubsection{Prior on flux}
\label{sect:PriorsFlux0}
A good flux estimator should be unbiased, but this goal is often
problematic.  The optimal estimators in the sense of decision theory,
i.e. those that minimise the expected loss/cost, are most often biased
and they combine the data with external information from ancillary
data sets.  PwSII thus includes two different sets of flux priors with
distinct goals.
\begin{itemize}
  \item \emph{Non-informative}. Our data model depends linearly on the
    source fluxes $A_j$ and is a particular case of the general linear
    model  \citep{BoxTiao}. Considering only a single source for
    simplicity (the solution for multiple sources is a mere repetition
    of this simpler case), one may show that the likelihood can be
    written in a form that makes it clear that the flux is in fact a
    location parameter:
  \begin{equation}
  \label{eq:PriorsLinearModel1}
   \mathcal{L}_{H_\mathrm{s}}(A_j) \propto \exp\left[- \frac{\sum_{\vect{\eta}} \mathcal{Q}_{jj}(\vect{\eta}) |\widetilde{\tau}(\vect{\eta};\widehat{\vect{a}}_i)|^2}{2} (A_j - \widehat{A}_j)^2 \right],
  \end{equation}
where $\widehat{A}_j$ is the MMF estimate of the flux
(\ref{eq:MaxLikeEstimate1}).  The same result could have been obtained
directly using formula (\ref{eq:PriorsRadiusNI0JR}). Thus, the prior
on the flux must be locally uniform:
\begin{equation}
\label{eq::PriorsFlux1}
\pi(A_j) \propto c,
\end{equation}
where $j$ indexes the source.  For a more general and rigorous
treatment see \cite{BoxTiao}.

  \item \emph{Informative}. Owing to the different statistical
    properties of point sources and SZ galaxy clusters, a different
    prior applies in each case.  For point sources, we adopt the flux
    prior first suggested by \cite{Argueso},
 \begin{equation}
 \label{eq::PriorsFluxGCauchy}
 \pi(A_j) = \Pr(A_j | A_0\,p\,\gamma) \propto \left[1 + \left(\frac{A_j}{A_0}\right)^{p}\right]^{-\frac{\gamma}{p}},
 \end{equation}
where $A_0$ is the `knee' flux, $p$ is some positive number and
$\gamma$ is the exponent controlling the shape of the power law for
fluxes much larger than the `knee'. This provides a good model for the
observed distribution of fluxes, fitting the de Zotti model almost
perfectly \citep{DeZottiSrcCounts}. Moreover, the distribution can be
properly normalised as required for evidence evaluation.  PwS
truncates the distribution faint tail and re-normalizes the remaining
range as result of the early selection effect (see
\ref{sect:PriorsModels}), a practice the proponents of the
distribution also followed. For galaxy clusters, the derivation of the
prior follows a different approach. The Planck Sky Model (PSM v1.6) \citep{AdamisSite}
was used to draw realistic simulations of the cluster populations
assuming a standard WMAP best-fit $\Lambda$CDM cosmology \citep{WMAP5}
and the Jenkins mass function \citep{JenkinsMassFunct}.  We found that
the fluxes in the sample cluster catalogues were quite well fitted by
a power law:
\begin{equation}
\label{eq::PriorsFluxClusterInf}
\pi(A_j) \propto A_j^{-\gamma}.
\end{equation}
To deal with the early selection threshold and to provide a properly
normalised distribution, once again a minimum and, this time, a maximum
flux also were assumed.
\end{itemize}

\subsubsection{Prior on size}
\label{sect:PriorsRadius}

\begin{itemize}
\item {\em Point sources}. Point sources are best modelled by imposing
  the prior $\pi(r) = \delta(r)$ on the `radius'.  This condition
  might, however, be too restrictive, since to simplify the
  implementation of the code and to make it faster, PwS assumes the
  instrument beams are circularly symmetric, which is only an
  approximation to the true beam shapes. Thus, even for point sources,
  allowing the source radius to vary over a small range of values allows
  a better fit between the template and the pixel intensities and
  consequently a higher likelihood ratio/SNR value. Thus, in both the
  informative and non-informative case, our preferred radius prior for
  point sources is
\begin{equation}
\label{eq:PriorsRadiusNI0}
\pi(r_j) =
\left\{
\begin{array}{ll}
  1/\Delta  &  r_j \leq \Delta  \\
  0  &  r_j > \Delta
\end{array}
\right.,
\end{equation}
where $\Delta \ll$ FWHM (the full-width-half-maximum of the beam).

\item {\em Galaxy clusters}.
Turning to galaxy clusters, a significant fraction of the clusters
Planck will detect will be unresolved, and thus appear as point
sources with a distinctive spectral signature. In many cases, however,
galaxy clusters are large enough to be mapped as extended objects and
a parameter controlling the scale of the cluster profile, the radius,
needs to be included.  The informative prior on the radius was derived
using the same procedure as in section~\ref{sect:PriorsFlux0} and
an exponential law
\begin{equation}
\label{eq:PriorsRadiusNI4}
\pi(r_j) \propto \exp\left(-\frac{r_j}{\ell} \right),
\end{equation}
was found to fit the simulated catalogues very well.
We truncate the distribution outside a minimum and maximum radius.

The non-informative prior follows a different law from that expected
from the cosmological models. Our model for an individual source is
the convolution of the source profile with the beam PSF.  The radius
parameter $r'_s$ that scales the resulting shape is a `hybrid'
parameter, as it shifts and scales the likelihood at the same time
(Jaynes, ch. 12). After applying the Jeffreys rule, the
non-informative prior on $r'_s$ reads:
\begin{equation}
\label{eq:PriorsRadiusNI1}
\pi(r'_s) \propto \frac{1}{r'^2_s}.
\end{equation}
Assuming that either the profile or the beam have centroids at the origin and the profile is a scaling profile $\tau(r/r_s)$ then,
\begin{equation}
\label{eq:PriorsRadiusNI2}
r'_s = \sqrt{B^2 + \kappa^2 ~ r^2},
\end{equation}
where $B^2$ is a constant known as the function variance of the beam \citep{Bracewell} and $\kappa^2$ is another dimensionless constant, the variance of the dimensionless variable $r/r_s$ over the profile.
The non-informative prior for the radius parameter then reads:
\begin{equation}
\label{eq:PriorsRadiusNI3}
\pi(r) \propto \frac{r}{\left(\mathcal{B}^2 + r^2\right)^{\frac{3}{2}}},
\end{equation}
where $\mathcal{B} = B / \kappa$.
For the general case $B^2$, the variance of the beam, should be replaced
by the variance of $\sqrt{\mathcal{P}_j(\vect{\eta})}$.
For unresolved
objects, narrow clusters with radii smaller than the beam size, the
prior grows linearly with $r$. For well resolved objects, $r \gg
\mathcal{B}$, the prior decreases proportionally to $r^{-2}$.
\end{itemize}

\subsubsection{Prior on spectral parameters}
\label{sect:PriorsSpectralInd}
There is an extensive literature on the distribution laws of radio
source spectral indexes : \cite{DeZottiSpecIndex},
\citep[][d,e]{PlnkRadioSpecIndex1}.
In general Gaussian distributions, or Gaussian mixtures with two
modes, fit the available data reasonably well.  However, the most
interesting sources are exactly those that do not follow the canonical
laws of emission. To avoid narrowing the range of possible
alternatives too much, uniform priors are probably better choices
unless we choose to target a very specific family. The same holds for
dusty galaxies.

By applying our standard procedure, the non-informative prior on the
spectral parameters reads
\begin{equation}
\label{eq:PriorsSpectralJeffreys}
\pi(\alpha_j) \propto \sqrt{\sum_\nu \left(\frac{\partial S_{\nu}(\alpha_j)}{\partial \alpha_j} \right)^2},
\end{equation}
where the sum extends over all frequency channels.

\subsubsection{Prior on the models}
\label{sect:PriorsModels}
The prior ratio $\Pr(H_{1})/\Pr(H_{0})$ on the models is often
neglected (i.e. assumed to equal unity), but plays a very important
role in the PwS detection criterion. To give a proper account of its
nature, let us imagine the simplest possible detection problem, where
we know in advance all the true values of the parameters that define
an object, which translates into delta-functions priors.
Substituting this condition into (\ref{eq:AveLossCatMak1}) and
making use of (\ref{eq:MaxLikeEstimate2}), we obtain the following
inequality:
\begin{equation}
\label{eq:PriorsSingleH0H11}
\mbox{SNR} \overset{H_1}{\underset{H_0}{\gtrless}} \sqrt{2~\left[\xi + \ln\left(\frac{\Pr(H_0)}{\Pr(H_1)} \right)\right]}.
\end{equation}
One may interpret the term $\ln\left(\frac{\Pr(H_0)}{\Pr(H_1)}
\right)$ as an extra `barrier' added to the detection threshold
because we are expecting more fake objects than the objects of
interest, due to background fluctuations.

We saw earlier that, when an object is present, a local maximum in the
likelihood is always present in the position parameter sub-space. This
condition immediately implies that only likelihood maxima need be
analysed. Nonetheless, one expects other likelihood maxima to occur as
a result of background fluctuation `conspiracies'.  Assuming Poisson
statistics for the number of sources and the number of likelihood
maxima resulting from the background fluctuations, then the ratio of
the probabilities is given by:
\begin{equation}
\label{eq:PriorsSingleH0H12}
\frac{\Pr(H_1 | N_s)}{\Pr(H_0| N_s)} = \left(\frac{\lambda_1}{\lambda_0}\right)^{N_s}
\end{equation}
where $\lambda_0$ is the expected number of maxima per unit area
resulting from background fluctuations above the minimum limit of
detection of the experiment, and $\lambda_1$ the expected number density
of sources above the same limit.

If only background is present, the density of maxima, $\lambda_0$,
resulting from the filtering procedure that creates the likelihood
manifold can be estimated using the 2D Rice formula:
\begin{equation}
\label{eq:PriorsLambda0Rice0}
n_{b}(\nu,\kappa,\epsilon)=\frac{8\sqrt{3}\tilde{n}_{b}}{\pi\sqrt{1-\rho^{2}}}\,\epsilon(\kappa^{2}-4\epsilon^{2})\,
e^{-\frac{1}{2}\nu^{2}-4\epsilon^{2}-\frac{(\kappa-\rho\nu)^{2}}{2(1-\rho^{2})}},
\end{equation}
where $\nu \equiv A/\sigma$ is the `normalised peak amplitude',
$\kappa$ the `normalised curvature', $\epsilon$ the `normalised
shear', and $\rho = \sigma_1^2/(\sigma_0 \sigma_2)$, with
$\sigma_{n}^{2} = (2\pi)^{1+2n} \int^\infty_0 \eta^{1+2n} |{\cal
  P}(\eta)|^2 ~d\eta$ \citep{RicePeaks}.  Marginalizing over all
parameters we obtain the expected density of maxima of a Gaussian
filtered field, which reads
\begin{equation}
\label{eq:PriorsLambda0Rice1}
\tilde{n}_{b}=\frac{\sigma_{2}^{2}}{8\pi\sqrt{3}\sigma_{1}^{2}}.
\end{equation}

One is not interested, however, in all peaks, but only on those above
a certain level $\nu_0$, since PwS pre-selects the putative detections
by imposing a minimum SNR level before attempting the evidence
evaluation. The main reason for adopting this early selection is
computational efficiency. The SNR alone provides a good proxy (see
formula \ref{eq:MaxLikeEstimate2}) for deciding whether a candidate
peak is the result of the presence of a source or just a background
fluctuation.  Moreover, low SNR peaks tend to be `badly-shaped' making
the sampler very inefficient and resulting in a very large fraction of
the samples being rejected. To make the things even worse, in most
cases, these peaks themselves end up being rejected as objects.

The applied flux cut must be taken into consideration to evaluate the
correct expected number counts, which define the prior $\Pr(H_1)$ as
well. Thus, $\lambda_0$ will read:
\begin{equation}
\label{eq:PriorsLambda0Rice3}
\lambda_0 = \int^\infty_{\nu_{0}} n_{b}(\nu) d\nu,
\end{equation}
where $n_{b}(\nu)$ is given by:
\begin{equation}
\label{eq:PriorsLambda0Rice2}
\begin{array}{r}
n_{b}(\nu)=\frac{\tilde{n}_{b}\,\sqrt{6}}{2\sqrt{\pi}\rho_{1}}\left\{ \left(1+{\scriptstyle \mbox{erf}}\left(\frac{\rho}{\rho_{1}\rho_{2}}\nu\right)\right)e^{-\nu^{2}\left(\frac{1}{2}+\left(\frac{\rho}{\rho_{2}}\right)^{2}\right)}\left(\frac{\rho}{\rho_{2}}\right)\:+\right.\\
\left(1+{\scriptstyle \mbox{erf}}\left(\frac{\rho}{\rho_{1}}\nu\right)\right)e^{-\frac{\nu^{2}}{2}}(\nu^{2}-1)\rho^{2}\rho_{1}\:+\\
\left.\frac{\nu
e^{-\nu^{2}\left(\frac{1}{2}+\left(\frac{\rho}{\rho_{1}}\right)^{2}\right)}}{\sqrt{\pi}}\rho\rho_{1}^{2}\,\right\},
\end{array}
\end{equation}
where $\rho_{1}=\sqrt{2\left(1-\rho^{2}\right)}$ and
$\rho_{2}=\sqrt{2\left(\frac{3}{2}-\rho^{2}\right)}$. The expected
number count of targeted objects above a certain flux threshold $S$,
$\lambda_1 \equiv \left<N(>S)\right>$, may be easily derived from
their differential counts.

Now a distinction must be made because the dominant type of
extra-galactic point sources in Planck maps are galaxies which, in
principle, do not follow the same statistics as the galaxy clusters.
From general cosmological assumptions it is possible to derive that
the expected differential counts for a certain population type of
galaxies per flux interval at a certain frequency always follow a
power law: $dN_{\phi}/dS = A_\phi~S^{-b}$ \citep{DeZottiSrcCounts}.
For clusters of galaxies, however, one must instead use a realistic
set of simulations, such as the `\emph{Planck Sky Model}' (PSM v1.6)
\citep{AdamisSite}. Using a properly normalised mass-function
\citep{JenkinsMassFunct}, one finds that a power law also fits quite
well the expect number counts of clusters above a certain threshold.
So, in either case, point sources or clusters, $\lambda_1$ may be written as:
\begin{equation}
\label{eq:PriorsLambda1}
\lambda_1 = N(>S_0) =  \int^\infty_{S_0} \frac{dN_{\phi}}{dS} dS = A_\phi~(1 - b)^{-1}~S_0^{1-b},\, \, b \neq 1,
\end{equation}
where we keep the parameters $\{A_\phi, b\}$ free.
These parameters are usually provided by the user to target a specific type of object and/or instrumental setup.
\section{Object detection strategy}
\label{sect:ObjectDetectStrat}
So far we have only developed the logic and probabilistic
underpinnings of PwS.  It is now time to bring all the pieces together
into a consistent strategy for the detection and characterisation of
discrete objects.  Our aim is to construct a robust, controlled, and
predictable algorithm.  Some caveats will be identified and
solutions suggested, always justified within the framework presented above.

\subsection{The single object approach}
\label{sect:ObjStratDetection}
Let us return to formula (\ref{eq:AveLossCatMak1}).  At a first look,
the evaluation of (\ref{eq:AveLossCatMak1}) seems quite a daunting
task. In order to apply the full Bayesian approach, many complex
integrals, over a very high dimensional volume (at least $4 \times
N_s$), need to be evaluated.\footnote{Even when working with one small
  patch at a time, seldom $N_s$ \, is smaller than 4.}.  Clearly a brute
force method is not efficient and perhaps not possible, even with the
massive computing resources generally available.

To find an effective solution, we begin by making two important
assumptions: (i) the objects of interest are `well separated', so that
(\ref{eq:WellSeparetedObjects}) holds; and (ii) all variables
pertaining to each individual source are mutually independent, which
has already been implicity assumed throughout the exposition of our
inferencial infrastructure.

These conditions allow us to separate the integrals associated with
each source. This is a very important simplification because it is now
possible to deal with each source independently, one at a time. This
is the `single object approach' \citep{mikecharlie} and replaces a
single $N_{\mathrm{param}} \times N_s$-dimensional integral with a
sequence of $N_s$ integrals, each of dimension $N_{\mathrm{param}}$.

The complete likelihood expression may now be replaced by the much
simpler `single source' form.  Although, we should exercise some care
in defining the limits of integration in position space, since no
significant likelihood mass can be shared among position integration
domains. Apparently, this requirement creates such a wealth of
complexity to the integral evaluation that the single source approach
might at first be considered a poor choice.  Fortunately, the method
PwS uses to evaluate the evidence integrals automatically enforces
this rule if the fields are not too crowded (see section
\ref{sect:ObjStratEvaluation}).
%\footnote{On average, for point sources, these need to be separated by
%  2 times the distance the $ASCE$ multiplied by the the source flux
%  takes to reach values much lower than the $rms$ of the
%  background. This can be a very large distance, much larger than the
%  beam FWHM if the background is strongly correlated as in the case of
%  Planck and the source is very bright. For extended objects this
%  distance increases as we need to convolve the $ASCE$ with the source
%  profile.}

Under our assumptions, the odds of the model $H_1$ (for a given source type),
given $N_s$ such sources, reads
\begin{equation}
\label{eq:ObjectDetect}
\frac{\Pr(H_1| \vect{d},N_s)}{\Pr(H_0| \vect{d},
  N_s)}=(N_\mathrm{pix}\Delta_\mathrm{p})^{-N_s}
e^{-\lambda_1} \frac{\lambda^{N_s}_1}{N_s !} \left(\frac{\lambda_1}{\lambda_0}\right)^{N_s} \prod_{j=1}^{N_s} \mathcal{Z}_{1j},
\end{equation}
where we have defined the `partial evidence' for each individual source as
\begin{equation}
\mathcal{Z}_{1j} \equiv \int
\frac{\mathcal{L}_1(\vect{\Theta}_j)}{\mathcal{L}_0}
\pi(\vect{\Theta}_j)\,d\vect{\Theta}_j.
\end{equation}
Taking logarithms and rearranging, one finds
\begin{equation}
\label{eq:ObjectDetectLn}
\ln \left[\frac{\Pr(H_1 | \vect{d}, N_s)}{\Pr(H_0 | \vect{d}, N_s)}\right]
= \sum^{N_s}_{j=1} \, \ln(\mathcal{Z}_{1j}) - N_sP_s,
\end{equation}
where we have defined the `penalty per source' $P_s$ as
\begin{equation}
P_s \equiv \ln \Lambda^{-1}_s  + \ln\left(\frac{\lambda_0}{\lambda_1}\right) + \frac{1}{N_s} \left[\lambda_1 + \ln N_s! \right].
\end{equation}
Thus, the total $\ln(\mbox{odds})$ for a single patch is the sum of
the partial $\ln(\mbox{evidence})$ for each source, plus an extra
global penalty term that contributes, in the majority of the cases,
negatively to the final balance and does not depend on any particular
source, but exclusively on the ensemble properties.

The most robust source catalogue is that which maximises the
$\ln(\mbox{odds})$ in (\ref{eq:ObjectDetectLn}), but we do not know
the value $N_s$. Moreover, we have not yet addressed how many or which
aspirant detections will be finally selected for inclusion in the
catalogue. Nonetheless, the expression (\ref{eq:ObjectDetectLn}) is a
sum, so its maximum value is reached when only the positive terms are
included. Thus, one possible procedure to select the optimal set of
sources is as follow:
\begin{enumerate}
  \item evaluate each partial $\ln$(evidence);
  \item sort them in descending order;
  \item to each line add the global $P_s$ term, with $N_s=1$ for the first line, $N_s=2$ for the next line, etc.\;
  \item starting from the line with the lowest partial
    $\ln$(evidence), move up throwing away all lines for which the sum of the
    above contributions is negative.
\end{enumerate}
The `proto-catalogue' is now made and $N_s$ found.

We are not finished yet, however, because we have only selected the
set of detections that maximises the odds. Other constraints may yet
apply.  For instance, we may impose a threshold per line different
from zero as result of the loss criteria or, as we shall see, a
prescribed contamination for the catalogue. Moreover, the $P_s$ terms
added to each line have different values, but only their sum has a
well-defined meaning, which applies to the full catalogue taken as a
whole.  Therefore, we next average the $P_s$ over the proto-catalogue
and add it to each partial $\ln$(evidence) to obtain the $\ln$(odds)
for each object
\begin{equation}
\label{eq:LogOddsPO}
\ln(\mbox{odds})_j \equiv
\left[\frac{\Pr(H_1 | \vect{d})}{\Pr(H_0 | \vect{d})}\right]_j
= \ln(\mathcal{Z}_{1j}) + \langle P_s\rangle.
\end{equation}
This quantity has a pivotal role in catalogue making (see section \ref{sect:ObjStratCatMaking}).
\subsection{Evaluation of the odds ratio}
\label{sect:ObjStratEvaluation}
Even using the simplified form of the likelihood assumed in the
single-object approach, a `brute force' evaluation of the resulting
evidence integrals is still not feasible. One must instead use a Monte
Carlo approach to the numerical integration. Evidence integrals are
usually evaluated using Markov Chain Monte Carlo (MCMC) methods and
thermodynamic integration. Such methods can fail, however, when the
posterior distribution is very complex, possessing multiple narrow
modes\footnote{At least one central maximum per source plus other
  secondary maxima around the central higher peaks \citep{PwSI}.} that
are widely separated. We therefore instead use `nested sampling'
\citep{SiviaG}, which is much more efficient, although not without its
difficulties. \cite{MultiNest} have developed a very efficient
implementation of the nested sampling algorithm, called `MultiNest',
which is capable of exploring high-dimensional multimodal posteriors.
Nonetheless, MultiNest is designed to be a general sampling and
evidence evaluation tool and it is not particularly tuned for
Planck.

In the interest of speed, PwS instead tries to take full advantage of
the properties of the astronomical data sets.  As already stated (see
section \ref{sect:Priors}), if our model explains the data well then
the likelihood should peak steeply around the parameter true values,
decay very rapidly to zero, and have most of its mass concentrated
around the maxima vicinities. Thus, if one can first find the
likelihood maxima, then one does not need a sophisticated multimodal
sampling algorithm like MultiNest. A much simpler nested sampling
scheme such as \cite{SingleNestSampl} would perform equally well.  Moreover,
reasonably high SNR maxima develop `well-shaped' peaks, in the sense
they are close to Gaussian, rendering the sampling highly efficient.
Two other significant advantages are: (i) we can reduce our data set
to a small neighbourhood enclosing the maxima, so that only a very
small number of pixels close to the maxima contribute appreciably to
the evidence value; and (ii) a much reduced parameter volume allows
the same number of `live points' to deliver a considerably higher
accuracy on the evidence value, since they do not split among the
several posterior peaks. This is the approach adopted in PwS, which we
now outline in more detail.

\subsubsection{Locating the likelihood maxima}
\label{sect:ObjStratFindingLikeMax}

Our first goal is to find the likelihood maxima. For illustration, let
us focus on the example of galaxy clusters, each of which is described
by $4$ parameters: $\{X, Y, S, R\}$.  An efficient 4-dimensional
minimiser implementation is straightforward and immediately available
\citep{NR}.  However, our manifold has many maxima and we need to check
all of them, otherwise we might lose some sources.

One possibility would be to follow the approach used in PwS I, where
the Brent line minimiser was 'enhanced' with an ancillary step to
allow it to `tunnel' from one minimum to the next one using a scheme
closely related with the equivalent quantum mechanical effect.  To
increase the effectiveness of the procedure, PwS I started a Powell
minimization chain (hence the name `PowellSnakes') in many different
locations of the manifold in an attempt to find all the maxima. It
should be remembered, however, that the likelihood only exhibits
multiple maxima in the position sub-space; the other sub-spaces are
`well behaved'. Moreover, the likelihood in the position sub-space is
merely the MMF filtered field. We therefore instead use a brute force
peak finding algorithm that scans all pixels in this subspace, which is
very easy to implement and almost instantaneous. Then, after
collecting a list of peak positions, we start a 4-dimensional
PowellSnakes optimisation at each such location to find the
maximum-likelihood parameters for that particular peak.

A subtlety does arise in this approach, however, since to obtain the
MMF filtered field, one needs to assume a size $R$ for the objects to
define the filter. Since we expect different clusters to have
different radii, we might lose some peaks because of the mismatch
between the true value of the cluster radius and that used in the
filtering template.  A simple solution would be that suggested by the
MMF authors: apply the filter repeatedly using a different radius each
time. Although practical, this is, however, not the most efficient
approach. Fortunately, if the instrument beams and the sources possess
reflection symmetries in both axes, then one can show that the Fisher
matrix at each likelihood peak is block-diagonal (assuming the
likelihood (\ref{eq:LikeFilterComplete}) and using the single-source
approach assumption (\ref{eq:WellSeparetedObjects})), such that there
is no correlation between the position subspace and the other
parameters (flux and size) of the cluster. This has two important
consequences: (i) regardless of the radius used to construct the
filter, a likelihood peak will always be present at the location
source and its position will not change positions as the filter scale
varies; (ii) we do not need to perform a full 4-dimensional
maximization but can (at least) separate the position variables from
all others, which brings a tremendous simplification to the problem of
finding the likelihood maxima. Thus, we can indeed start by finding
the maxima in the position subspace using a brute force
`check-all-pixels' approach and then, after pinpointing the
position of the source, search the remaining sub-spaces associated
with the other variables.

A couple of final comments on this approach are worth making. First,
it is well known that matched filters are excellent at finding and
locating sources, but not as good at estimating fluxes.  If the beam
shape/size is not completely known but symmetric, even when building
up a filter with the wrong beam geometry, the filter will correctly
recover the positions of the objects. In general, however, the element
in the Fisher matrix corresponding to the correlation between the
radius and the flux of an object is non-zero.  Therefore, if the
filter is assembled using wrong beam parameters, bias in the flux
estimates must be expected. Second, and perhaps more subtle, is that
the symmetries of the Fisher matrix only hold on average.  Thus, for
each individual peak some residual correlation between the position
and the other variables is expected.  According to our current
accumulated experience, however, this correlation is usually very
small. Nonetheless, PwS still includes the option to use the peaks
positions obtained from the MMF filtered fields just as initial hints
for a full N-dimensional Powell minimisation.

\subsubsection{Exploring the posterior distribution}
\label{sect:ObjStratAndThePriors}

Our initial step provides the ML estimates and the SNR of each
detection candidates. This has a very useful side effect, since we do
not need to explore the posterior distribution around all the maxima
we find. Only a much smaller sub-set is chosen based on an SNR
threshold. This SNR threshold should be low enough not to reject any
substantial fraction of peaks associated with true detections and high
enough to make the selected sample contain a large percentage of true
sources and to include most `well-shaped' maxima. This shorter list is
then sorted in descending order of SNR and one-by-one the maxima are
sent to the nested sampler, which returns an evidence estimate and a
set of weighted samples that we use to model the full joint posterior
distribution. From these samples we can compute any parameter
estimate, draw joint distribution surfaces, predict HPDs intervals of
any content over the marginalized distributions to infer the parameter
uncertainties, etc., as in the example presented in \citep[][c - fig. 9]{PlnkEarlySZ}.
The current released implementation of PwS (v3.6) computes the maximum-likelihood, the expected value over the
posterior estimates and $1 \sigma$ error bars.\footnote{The
  maximum-likelihood estimates from the maximization step are updated,
  if necessary, during the sampling phase.}

\subsection{Non-gaussianity of the background}
\label{sect:ObjStratNonGaussianity}

It is clear that our model of the observations, like any
  model, is only an approximation to the real data. This is true both
  for our model of the discrete objects and for our model of the
  background. For the latter, it is clear that the background emission
  in real observations is neither Gaussian nor statistically
  homogeneous. Regarding non-Gaussianity, we do not mean that of a
  primordial origin, which, if exists, would have an insignificant
  effect in our analysis. We are instead alluding to the
  non-Gaussianity induced by the Galactic emission components, the
  confusion noise created by the sources below the detection
  threshold, the instrumental noise artefacts coming from the
  incomplete removal of the cosmic rays glitches and, of course, a
  wealth of other possible sources.

Many authors simply ignore this issue and many others dismiss its
importance. A very strong argument, used many times, is that despite
the sky emission being admittedly non-Gaussian, the effect of the
finite PSF of beams will combine many different sky locations into a
single pixel. In addition, signal de-noising procedures further
combine more samples. Some authors then appeal to the Central Limit
Theorem (CLT) to claim that non-Gaussian effects in the final data
must be completely negligible.

This argument seems particularly appealing, but a deeper analysis of
the CLT shows that, in our particular problem, namely detection and
separation of two signals, the effects of the CLT are not as important
as those authors claim.  Formally, the CLT only applies when $N
\rightarrow \infty$, where $N$ is the number of random deviates in the
sum. For finite $N$, the CLT only guarantees the Gaussian
approximation is good for `\emph{a region around the mode}'
\citep{RiskAnalys}. The size of this Gaussian region grows very
slowly. In the worst case, the distributions of the individual
deviates are skewed and have `fat tails'. Let us focus on a real
example: the Galactic emission.  If the spectral brightness
distribution follows a power law with a finite second moment, to
guarantee the field has physical behaviour ($ \equiv $ finite energy),
the normalised central Gaussian region, $|u|$, only grows very slowly
with $N$:
\begin{equation}
\label{eq:NonGaussianRegion}
    |u| \ll u_0 \propto \sqrt{\ln N}
\end{equation}
where $u_0$ is the tail lower boundary. This means that the sum must have
more than $1000$ terms to make the Gaussian approximation acceptable
up to about $|u| \sim 2.6$.
In detection problems, where we want to separate the maxima created by
the sources from the background fluctuations, we are dealing all the
time with the background distribution upper tail:
\begin{equation}
\label{eq:NonGaussianUpperTail}
\mathcal{P}_{\underset{u_0}{>}} \equiv  \int^{\infty}_{u_0} \Pr(u) ~ du.
\end{equation}
If the background field intensity distribution follows a power law:
$\Pr({I_\nu}) \propto {{I_\nu}}^{-\mu}$, with $\mu > 2$, to
guarantee its energy is finite, then the probability that a sum of $N$
deviates falls into the upper tail region of the sum normalised
distribution is:
\begin{equation}
\label{eq:NonGaussianProb}
\mathcal{P}_{\underset{u_0}{>}} \propto \frac{1}{N^{\mu/2 - 1} \ln^{\mu/2} N}.
\end{equation}
This is a very serious problem. Object detection methodologies are
designed typically to suppress the background and amplify what does
not fit its model. The non-gaussianity component is not part of our
background model, so its effect on the detection process is doubly
pernicious: not only it is not removed, it is amplified.

There seem to be only two ways of circumventing this problem: (i) to
include the non-Gaussian effects in the statistical models; and (ii)
to manipulate and add as much data as possible to make it more
Gaussian.  Owing to the complexity of Planck data it is almost
impossible to give a proper account of the non-Gaussian effects
without making the problem unsolvable. So, a workable solution must
necessarily combine as much data as possible, and then analyse the
outcome. The only possible way of doing this is to use multi-channel
analysis all the time.

Our own experience corroborates this view.  The SNR values
of the PwS selected detections and the thresholds the frequenstist
methods normally employed ($\gtrsim 4.0$), are much higher than what
would be expected according to the purity levels of the catalogues if
the statistics were purely Gaussian.  Although, the channels with the
largest beams, where each pixel is the result of a much higher number
of different contributions, do indeed have detection thresholds lower
and closer to those expected from the Gaussian theory.  A good
practical example of how the multi-channel processing can help the
reduction of the impact of the non-Gaussian distributions on the
detection process is the recovery of the SZ signal
\citep{MelinChallenge}.

Owing to the residual non-Gaussianity left in the background,
especially close to the Galactic plane, we should now expect a higher
number of background fluctuations reaching above the evidence
threshold level than those predicted by the Gaussian model.  So,
eventually, we need to correct the prior on the models:
$\frac{\Pr(H_{1})}{\Pr(H_{0})}$, as this prior was derived assuming
that the background had purely Gaussian statistics.  The simplest way,
we believe, is just to count the total number of fluctuations above
the SNR threshold adopted, before embarking on the evaluation of
the evidence.
In particular, one should compare this number with what
  would be expected from the Gaussian model plus the predicted source
  counts above the SNR threshold, then take the larger
  quantity. Denoting this value by $T$, a corrected estimate of
  $\lambda_0$ (see formula \ref{eq:PriorsSingleH0H12}) would read
\begin{equation}
\label{eq:PriorModelCorrection}
\lambda_0 \simeq  T - \lambda_1.
\end{equation}
This very simple `trick' provides a first-order correction
to the effects of background non-Gaussianity.

\subsection{Statistical inhomogeneity of the background}
\label{sect:ObjStratBackAccidents}

Real observations will also inevitably exhibit some
  statistical inhomogeneity of the background, in contradiction to our
  assumed model.  Consequently, the conditions of optimality derived
  therefrom no longer hold. This can lead to a number of difficulties
  in detecting and characterising discrete objects, particularly in
  regions of the sky that contain bright, very inhomogeneous and
  anisotropic backgrounds. Indeed, this general expectation has been
  borne out in applying earlier versions of PwS to detailed
  simulations of Planck observations \citep[PSM 1.6,][]{AdamisSite}.
  In particular, the presence of bright diffuse Galactic dust emission was found to lead to the PwS SZ catalogue (in
  common with catalogues produced by other methods, such as MMF)
  containing bright spurious detections. Hence one did not obtain a
  regular cumulative purity curve that slowly approaches unity as the
  $\ln$(evidence), or the SNR, increases \citep{MelinChallenge}, in
  contradiction to what would be expected from theory if our model
  explained the data properly.

Indeed, the detection of SZ galaxy clusters highlights further
problems.  Again in the analysis of Planck simulations using previous
versions of PwS, one finds that bright spurious SZ signals are not only
concentrated in complex background regions, with a fraction of
the bright spurious detections spread all across the sky. By
cross-correlating the resulting SZ catalogues with ancillary point
source data sets, one finds that bright spurious cluster detections
matched bright point source locations. In our preliminary attempts to
address this problem, we therefore first performed a point source
extraction step and subsequently subtracted/masked the best-fit point
source profiles in the maps. This pre-processing step greatly helped
in reducing the number of spurious detections, especially those with
very high evidence values. Another approach has been suggested by the
Planck WG5 team, namely the `$\chi^2$ test' \citep[][c]{PlnkEarlySZ}.
This performed very well, although, once more, there is no easy way to
choose a robust acceptance/rejection threshold for the test. Another
difficulty occurs when extracting the SZ effect at each individual
channel. The SNR was usually so low that the measurements ended up
being quite noisy.

Can we do any better using Bayesian logic? The apparent failure of
the `best' test can be immediately explained using the main
Bayesian decision equation, equation (\ref{eq:AveLossCatMak1}).
Our decision criterion is based on the $\ln(\mbox{odds})$, namely
\begin{equation}
\label{eq:BrightSpuriousDeci}
    \ln \left[ \frac{\Pr(H_1 | \vect{d})}{\Pr(H_0 | \vect{d})} \right].
\end{equation}
The problem comes from the denominator $\Pr(H_0 | \vect{d})$. When we
find a point source, its probability of being a cluster, $\Pr(H_1 |
\vect{d})$, is very low, but the probability of those pixels being
part of the background, $\Pr(H_0 | \vect{d})$, is also very low,
because point sources do not fit our model of the background either.
We have already mentioned that the binary model is too simple to
handle realistic astronomical situations. To secure the optimality of
our methodology we must ensure that the data is well described by our
model, and employ a multi-model approach, as described in
Section~\ref{sect:ObjStratMultiModel} below.

\subsection{The solution: multi-model, multi-frequency detection}
\label{sect:ObjStratMultiModel}

For the reasons outlined above, we believe that a deeper and purer
catalogue can only be obtained through multi-frequency analysis, which
cannot be pursued without assuming some spectral signature. An
excellent example of the power of such an approach is provided by the
detection of SZ clusters.  Despite the SZ signal being sub-dominant on
all Planck channels (the signal level is below that of the
background), an optimal combination of the different frequencies can
boost these extremely faint signals to the point where one can
now build reliable catalogues of many hundreds of such objects

We have also demonstrated above that our simple binary decision making
approach is too na\"{\i}ve to handle `real-life' situations.  The
introduction of a multi-model (more than two models) decision rule
cannot, however, be achieved simply by extending the binary case
(Jaynes, Ch. 3), although it is always possible to reduce the general
multi-model decision rule to a succession of binary ones.  We start by
choosing one of the hypothesis, say $H_0 \equiv$ `\emph{this maximum
  is a background fluctuation}', and making it the `\emph{null}' or
`\emph{reference hypothesis}'.  Then we iterate through all the
hypotheses associated with different source families and we compute
the $\varrho_i \equiv \mbox{odds}_i$:
\begin{equation}
\label{eq:MultiModelOdds_i}
\varrho_i \equiv \frac{\Pr(H_i|\vect{d})}{\Pr(H_{0}|\vect{d})}; \quad i \neq 0
\end{equation}
The optimal way of deciding between $M+1$ different hypothesis ($M$
source types plus the null hypothesis) is by evaluating the odds for
each type of source against the null hypothesis, pick up the largest
$\varrho_{i}$, which we denote by $\varrho_{i^\ast}$, and then check
for the following inequality
\begin{equation}
\frac{\varrho_{i^\ast}}{1 + \sum_{_{i \neq  i^\ast}} \varrho_i} \overset{H_{i^\ast}}{\underset{H_0}{\gtrless}} \xi,
\label{eq:BI_PostProbsRatioMult}
\end{equation}
where $\xi = L_{\text{spurious}} / L_{\text{miss}} $ is the ratio of
the losses when accepting a false positive (spurious) and when missing
a source. We are implicitly assuming that the penalty for choosing
wrongly in favour or against the most probable hypothesis,
$H_{i^\ast}$, is always $L_{\text{spurious}} ~ \text{or} ~
L_{\text{miss}}$ regardless of the true/alternative hypothesis and
there is no loss when choosing wrongly between any of the alternative
hypotheses.

A pivotal quantity in catalogue making, as we shall shortly see, is
the probability that a certain entry in the putative catalogue is a
spurious detection: $\Pr(\widetilde{H_{i^\ast}} | D)$.  Providing this
value is a unique capability of the Bayesian approach.  It is very
simple to show that, when extending the binary test to multiple
hypotheses, the probability of a spurious detection now reads (Jaynes,
Ch 3):
\begin{equation}
\label{eq:BI_PostProbsRatioMultProbSpur}
    \Pr(\widetilde{H_{i^\ast}} | \vect{d})
= \frac{1}{1 + \psi}, \quad\mbox{with}\,\,\, \psi \equiv \frac{\varrho_{i^\ast}}{1 + \zeta},\quad \zeta \equiv \sum_{i \neq i^\ast} \varrho_i.
\end{equation}
\subsection{Catalogue making}
\label{sect:ObjStratCatMaking}

The last step of PwS is to assemble the final catalogue from a list of
candidates. During this stage, PwS performs the following steps:
\begin{enumerate}
  \item maps flat sky patches back onto the sphere at the
the positions of the putative detections;
  \item applies a detection mask, if any;.
  \item merges multiple detections of the same source
obtained in different patches into a single candidate detection; and
  \item makes the final catalogue by rejecting those lines that do
    not meet the pre-established criterion of purity or loss.
\end{enumerate}
The last step is critical to the success of our methodology.  We
already gave some indication in
Section~\ref{sect:DecisionTheoryCatMaking} about how to address the
difficult task of selecting a sub-set of detections from our initial
list of candidates. If the selection criterion is based on losses,
then we just need to trim the `proto-catalogue' further by applying
the decision rule (\ref{eq:BI_PostProbsRatioMult}).  But, as we
mentioned previously, it is much more common in astronomy to require a
catalogue to have an expected contamination ratio or that the
contamination does not exceed a prescribed value. We are now finally
in position to show how the Bayesian logic framework can give us
exactly that.

The number of false positives in a catalogue may be represented as a
sum of Bernoulli variables.  Assuming all catalogue entries are
statistically independent, then the sum of $N$ of those variables is
distributed as a Poisson--binomial distribution:
\begin{equation} \label{eq:PoissonBinDist}
\mu = \sum_{i=1}^{n} p_i, \quad \sigma^2 = \sum_{i=1}^{n} p_i (1 - p_i),
\end{equation}
where $p_i = \Pr_i(\widetilde{H_{i^\ast}} |\vect{d})$, is the probability of
source $i$ being a false positive.

Therefore, one way to proceed is as follows:
\begin{enumerate}
  \item sort the list of candidate detections in $\ln(\mbox{odds})$
    descending order ($p_i$ ascending order);
  \item  for each candidate, accumulate $p_i$ until $\mu$ (see formula \ref{eq:PoissonBinDist}) exceeds the prescribed contamination
$\alpha \equiv \mbox{(spurious detections)}/\mbox{(total lines in catalogue)}$ times the total number of lines already included; and
\item discard the last line.
\end{enumerate}
As $\mu$ is a sum of independent variables and $N$ is usually a large
number (hundreds), it is perfectly reasonable to assume the distribution
converges to a Gaussian as result of the CLT
\footnote{Note this time we are working around the distribution mode.}.
So, a good estimate of the number of spurious detections in the catalogue is
\begin{equation}
\label{eq:CatalMak1}
    \sum_{i=1}^{N} p_i \pm \sqrt{\sum_{i=1}^{N} p_i (1 - p_i)},
\end{equation}
and an estimate of the fraction of spurious detections in the
catalogue, $\alpha$, reads:
\begin{equation}
\label{eq:CatalMak1}
    \left( \widehat{\alpha} = \frac{\sum_{i=1}^{N} p_i}{N} \right) \pm \frac{\sqrt{\sum_{i=1}^{N} p_i (1 - p_i)}}{N}.
\end{equation}

A problem still remains, however, since our calculation of
$\Pr_i(\widetilde{H_{i^\ast}} | D)$ is only an approximation, although
we do have an estimate of the $\ln(\mbox{odds})$ evaluation
uncertainty (for a rigorous treatment see \cite{MultiNestUncert}). We
therefore need to introduce corrections into the above formulas to
account for the uncertainty on $p_i$. It is easy to verify
that, to a first approximation, the error on $p_i$, reads:
\begin{equation}
\label{eq:CatalMakErrEvid}
|\Delta p_i| \simeq \gamma p_i (1 - p_i)
\end{equation}
where the value of $\gamma$ is the average evidence evaluation
fractional error.  The corrected value of the catalogue's variance on
the number of spurious, $\sigma'^2$, is always less than:
\begin{equation}
\label{eq:CatalMakErrEvid1}
\sigma'^2 \lesssim (1 + \gamma) \sum_{i=1}^{n} p_i (1 - p_i),
\end{equation}
and the variance on $\mu$ reads:
\begin{equation}
\label{eq:CatalMakErrEvid2}
|\Delta\mu|^2 \simeq \gamma^2 \sum_{i=1}^{n} p^2_i (1 - p_i)^2 < \gamma \sum_{i=1}^{n} p_i (1 - p_i)
\end{equation}
Thus, we get the final expression of predicted contamination
of the catalogue by adding both contributions in quadrature:
\begin{equation}
\label{eq:CatalMak2}
    \left( \widehat{\alpha} = \frac{\sum_{i=1}^{N} p_i}{N} \right) \pm  \frac{\sqrt{1 + 2\gamma} \,\sqrt{\sum_{i=1}^{N} p_i (1 - p_i)}}{N}.
\end{equation}
The uncertainty on the contamination of the catalogue for commonly
accepted levels ($ \sim 10\%$), catalogue size ($ \gtrsim 1000$) and
$\gamma$ as large as $0.32$ (value taken from the extraction exercises
with Planck data), is always $\lesssim 1.2\%$.

Finally, we are now in position to answer the key,
  question all the frequentist methods must at some point face,
  \emph{``What threshold should one use for accepting the candidates
    for inclusion in the final catalogue?''}, although the question is
  no longer relevant in our Bayesian approach, since it is an output
  of our catalogue-making method, rather than an input.  The answer is
  just \emph{``the $\ln(\mbox{odds})$ estimate of the last line of the
    final catalogue}'', since the initial list of putative detections
  was sorted in descending order of $\ln(\mbox{odds})$ and all those
  with a higher or equal $\ln(\mbox{odds})$, and only those, were
  selected for inclusion.

\section{Implementation history}
\label{sect:Implementation}
%

%The versions that have been used in published data analyses so far are v3.1
%for the SZ Challenge \citep{MelinChallenge}, v2.01  for the lower
%frequency sources in
%the Planck ERCSC \citep{PlnkERCSC}and for all frequency channels in
%\citep{RochaCSI} and 3.6 in application to SZ cluster detection
%in the Planck ESZ sample \citep{PlnkEarlySZ}.

The data analysis philosophy and set of algorithms described in this
paper have not so far been fully implemented in a coded version of
PwS.  We are working towards this aim, and the release corresponding
to the full set of features described here will be PwS v4.0.
The versions that have been used in published data analyses so far are v1.5 and v3.1
for the SZ Challenge \citep{MelinChallenge}, v2.01  for the lower frequency point sources in
the Planck ERCSC \citep[][b]{PlnkERCSC} and for all frequency channels in the Compact Source Investigation workshop (CSI) \citep{RochaCSI}, v3.6 in application to SZ cluster detection in the Planck ESZ sample \citep[][c]{PlnkEarlySZ} and to characterise a single cluster parameters in a non-blind exercise \citep[][g]{Arnaud_266}.
It is worth noting that these versions
include a pre-processing tool specifically designed to convert data sets
distributed within the Planck collaboration into the format required by PwS.
The main tasks performed by this tool are:
%;
\begin{itemize}
  \item taking account of the masking and/or flagging of ill-observed
    pixels and contaminated regions;
  \item projecting the spherical maps into flat patches\footnote{The
    patch set usually contains about $~12,000 ~ 7.33^\circ \times
    7.33^\circ$ flat patches or $~3,000 ~ 14.66^\circ \times
    14.66^\circ$ instead};
  \item mapping of coordinates from the sphere into the patches and
    back;
  \item removal of multiple detections of the same source in different
    patches;
  \item assembly of the output catalogues into the required format.
\end{itemize}

The existing released versions of PwS differ from what will
be available in v4.0 mainly in the limitation to a binary model
selection step in determining when to accept a putative source detection and
to a non-parametrised frequency spectrum in multi-frequency
detection. The latter restriction meant that, while SZ cluster
detection could be carried out using all Planck frequencies
simultaneously, point source detections, in common with the other
methods available, were carried out for each frequency channel
separately. PwS v4.0 will aim at genuine multi-frequency and indeed
multi-model detection, using all the available data simultaneously.

\section{Conclusions}
\label{sect:Conclusions}
The Planck satellite, and many other modern cosmological, data sets
present completely new challenges for the detection and description of
compact objects. Two important traits of such observations are (i) low
or very low SNRs; and (ii) strongly correlated backgrounds with
typical scales similar to those of the objects being sought. These
attributes render traditional object detection methods sub-optimal,
since: (i) it is difficult to separate the sources from the
background fluctuations; and (ii) the uncertainties on derived
source parameters are important and traditional methods do not provide them.

A better strategy is to develop an object detection methodology from a
strong statistical foundation first.  The linear filtering family of
tools is the attempt by the orthodox frequentist school of probability
to overcome these limitations. The matched filter and all its
derivatives are based on the Neymann--Pearson likelihood ratio,
although their optimal performance is extremely dependent on the
choice of the acceptance/rejection threshold and on implementation
details. Despite of their widespread use, the actual practical designs
of these tools do not yet implement a sound framework to handle the
uncertainties on the parameter estimates.

Bayesian methods have the great advantage of providing a coherent
probability methodology with the option to include, in a completely
consistent way, all ancillary information. But probability theory by
itself only gives us a degree of belief. In order to produce a
catalogue, decisions must be made as well. Decision Theory is
unambiguous: $\ln \left[\frac{\Pr(H_1 | \vect{d})}{\Pr(H_0 |
    \vect{d})} \right]$ is the optimal decision tool (in the binary
case), although the binary model is manifestly not powerful enough to
handle a real data set. The necessary extension to a multi-model
foundation is mandatory for an operational and viable solution.

In PwS we have attempted to implement a fast, multiple model decision
rule based on the Bayesian $\ln(\mbox{odds})$ device.  To achieve our
goal we focused on taking advantage of the symmetries of the
multi-channel likelihood manifold to design an efficient, though
rigorous, exploration tool. Owing to its full, consistent probability
foundation, PwS can provide a sound and complete statistical
characterization of its results. Simultaneously, we can offer
effective solutions for the difficulties accompanying real data,
without compromising any of our goals.
\section{Acknowledgments}
P. Carvalho thanks all his colleagues at the
Astrophysics Group of Cavendish laboratory and the KICC, and his fellow
members of the Planck consortium for their help in completing PowellSnakes.
In particular, special thanks go to Farhan Feroz for his insightful
contributions and discussions. P. Carvalho is supported by a Portuguese
fellowship (ref: \emph{SFRH/BD/42366/2007}) from the Funda\c{c}\~{a}o
para a Ci\^{e}ncia e Tecnologia (FCT).\\
GR gratefully acknowledges support by the NASA Science Mission
Directorate via the US Planck Project. The research described in this
paper was partially carried out at the Jet propulsion Laboratory,
California Institute of Technology, under a contract with NASA.

\end{document}